\documentclass[intlimits,twoside,a4paper]{article}

\usepackage[cp1251]{inputenc}

\usepackage[eqsecnum]{cmpj3}

\issue{2022}{25}{3}{33602}
\doinumber{10.5488/CMP.25.33602}
\title[Effects of porous media on the phase behaviour of polypeptide solutions]%
{Effects of porous media on the phase behaviour of polypeptide solutions}
\author[V. I. Shmotolokha,  M. F. Holovko]{V. I. Shmotolokha\orcid{0000-0003-2729-4777}\thanks{Corresponding author: \email{shmotolokha@icmp.lviv.ua}.},
        M. F. Holovko\orcid{0000-0001-8114-5356}}
\address{
 Institute for Condensed Matter Physics, 1 Svientsitskii St., 79011 Lviv, Ukraine
}

\date{Received April 19, 2022, in final form July 13, 2022}

\begin{document}

\maketitle

\begin{abstract}
The generalized van der Waals equation for anisotropic fluids in porous media, proposed by the authors in previous works, is used to describe the effect of porous media on the phase behavior of polypeptide solutions. By introducing the temperature dependence for the depth of the potential well and the geometric parameters of the spherocylinder, the main features of phase behavior of the polypeptide poly ($\gamma$-benzyl-$L$-glutamate) (PBLG) in a solution of dimethylformamide, including the existence of two nematic phases, is reproduced. It is shown that the presence of a porous medium shifts the phase diagram to the region of lower densities and lower temperatures.
%
%
\printkeywords
%
\end{abstract}

\section{Introduction}


Physico-chemical research on the physics of polymers has long focused mainly on the study of linear flexible polymers, the flexibility of which depends on the chemical structure of the polymer, its chemical functional groups and environment \cite{1_Flory1953,2_Gennes1979}. Stiff and semi-stiff polymer systems have also attracted considerable attention in recent decades. Macromolecules of such polymers are quite stiff or contain stiff functional groups, as a result of which the polymer systems may exhibit liquid crystal properties \cite{3_Miller1978,4_EgMil2016}. As a result, polymer systems can be characterized by a wide range of phase transitions, which make their properties different from the properties of flexible polymers.

Liquid crystal orientation was first observed in biopolymers, in particular in rod-like viruses such as tobacco mosaic viruses \cite{5_Bawden36,6_Par76}. There are also a number of other biological polymers that exhibit liquid crystal properties. Among them are peptide compounds whose molecules consist of two or more residues of $\alpha$ -amino acids which are connected in an unbranched chain by a covalent peptide bond -C(O)-N(H)-. Depending on the number of amino acid residues in the chain, polymers belong to the family of polypeptides or proteins. Most polypeptide and protein molecules are fairly stiff rod-like molecules. In particular, proteins such as collagen, spectrin, myosin, actin and keratin have a stiff or semi-stiff rod-like structure, which is essential for their biological functions \cite{7_Tracy}.

In aqueous solutions, polymers are studied as polyelectrolytes. Therefore, along with the shape of molecules, electrostatic interactions also play an important role in them. The study of nonionic synthetic helix polypeptides in nonpolar organic solvents was initiated by \cite{8_ElAm1950,9_Rob1956}, which showed an important role of  nonspherical shape of molecules in the formation of the liquid crystal phase. Particular attention was paid to the study of the phase behavior of a solution of a polypeptide of poly ($\gamma$-benzyl-$L$-glutamate) (PBLG) in dimethylformamide (DMF), which is characterized by significant regions of nematic order stability and interesting phase coexistence of two liquid crystal states with noticeable concentrations of polypeptides. Key experimental data of the research of the phase behavior of PDLG solutions in DMF was obtained in the Miller group \cite{10_WeeMil1971,11_MilWuWeeSanRaiGo74,12_ZimWuMilMas83,13_RusMil1983,14_RusMil1984}, which use  various experimental techniques, such as proton nuclear magnetic resonance (NMR), polarization spectroscopy, differential scanning calorimetry (DSC) and viscometer. The phase behavior of the PBLG solution in DMF was also discussed in detail in \cite{15_McCarMilCairMor90,16_Ginzburg03}.

For the theoretical interpretation of the obtained phase behavior of the PBLG solution in DMF in the first works \cite{10_WeeMil1971,11_MilWuWeeSanRaiGo74}, the Flori lattice model was used \cite{17_Flory1956,18_Flory1956}. However, despite some qualitative agreement between theoretical and experimental data, at a discrete representation of the solution in a lattice consideration and unnatural interpretation of rod-like molecules, it is difficult to make a direct comparison between the lattice model and real macromolecules. A more realistic non-lattice model for describing the phase behavior of a PBLG solution in DMF was first proposed relatively recently in the Jackson group \cite{19_Wu}. In the proposed model, the PBLG solution is considered as a system of hard spherocylinders with attraction in the form of an anisotropic potential well. This model is described in the framework of the previously developed \cite{20_Franco-Melgar} approach, which combines the Onsager \cite{21_Onsager49} approach for describing rather long hard spherocylinders with the van der Waals approach that account for the attractive interactions \cite{22_YukhHol80}. To consider the change in the conformation of the macromolecule PBLG, the authors of \cite{19_Wu} introduced the temperature dependence of the parameters of the spherocylinder and the depth of the potential well with the appropriate fitting parameters. As a result, the authors were able to reproduce the experimental phase diagram quite well. However, to describe the orientational ordering, the authors used a trial function in the Onsager form \cite{21_Onsager49}, which significantly overestimates the orientational ordering \cite{23_Vroege92}.

It should be indicated that all description of isotropic-nematic transition in non-spherical particles since Onsager's theory \cite{21_Onsager49} can be considered as different variants of perturbative density functional theory type. In all cases, the free energy of the considered systems is formulated with some approximation as a functional of the orientation-dependent singlet distribution function $f(\Omega)$. The minimization of the free energy (or grand canonical potential) of the singlet distribution function leads to a non-linear integral equation for singlet distribution function, which can be solved numerically or by using the trial function with the coefficients, which can be found by minimizing the free energy. An example of such an approach was recently presented in \cite {23a_Nascimento}, where for a description of thermodynamics of anisotropic systems of particles with steric interactions there was used a formalism of functional density, which was founded by~\cite{36_Parsons79,37_Lee87} and then generalized in the theories by van der Waals.

A well-studied phase diagram of a solution of PBLG in DMF could work as a good base model for understanding the phase behavior in biopolymers. The next step should be to take into account the biological environment, which is often seen as a porous environment \cite{24_Vaf2010}. This is the purpose of the present work. The theoretical basis of the work is based on the works \cite{25_HolShmotPat, 26_HolShmot2015,27_HolShmot2020}, in which the generalization of  van der Waals equation on anisotropic fluids in porous media is given. In contrast to~\cite{20_Franco-Melgar}, the orientational ordering in the system is described through a corresponding integral equation for the singlet distribution function, which is found by minimizing the free energy functional of the system and its numerical solution on the basis of generalization \cite{28_HerBerWin84}.

\section{Theory}

As in \cite{19_Wu, 20_Franco-Melgar}, we consider a model of hard spherocylinders with length $L_{1}$ and diameter $D_{1}$ with attraction, which we choose to describe as an anisotropic potential well
\begin{align}
	u^{\text{attr}}\left(r,\Omega_{1},\Omega_{2}\right)&=\left\{\begin{array}{ll}
		-\left[\epsilon_{0}+\epsilon_{2}P_{2}(\cos { \gamma_{12}})\right],&
		\gamma_{1}D_{1}>r>\sigma\left(\Omega_{1},\Omega_{2},\Omega_{r}\right),\\
		0,& r< \sigma(\Omega_{1},\Omega_{2},\Omega_{r}),\,\,\, r>\gamma_{1}D_{1},
	\end{array}\right.
	\label{hol_smot37}
\end{align}
where $\epsilon_{0}$ and $\epsilon_{2}$ characterize  isotropic and  anisotropic parts of the attractive interaction respectively, $\gamma_{1}=1+{L_{1}}/{D_{1}}$, $P_{2}(\cos \gamma_{12})={1}/{2}(3\cos^{2} \gamma_{12}-1)$ is the second order Legendre polynomial of  relative orientation, $\gamma_{12}$ is the angle between two  principal axes of two spherocylinders. We hope that the reader will not confuse the angle $\gamma_{12}$ with notation $\gamma_{1}=1+{L_{1}}/{D_{1}}$.  $\sigma(\Omega_{1},\Omega_{2},\Omega_{r})$ is the orientation-dependent contact distance of two particles, $\Omega_{1}$ and $\Omega_{2}$ are orientations of two particles 1 and 2,    $\Omega=(\theta,\phi)$ denotes the orientations of particles, which are defined by the polar angles $\theta$ and $\phi$. $\Omega_{r}$  is defined by the angles between the fixed system of coordinates and the system of coordinates,  which moved with two considered particles. In terms of $\sigma(\Omega_{1},\Omega_{2},\Omega_{r})$, the repulsive part of the interaction $u^{\text{rep}}(r_{12},\Omega_{1},\Omega_{2})$ for hard particles can be represented in the form

\begin{align}
	u^{\text{rep}}(r_{12},\Omega_{1},\Omega_{2})
	=\left\{\begin{array}{ll}
		\infty,\quad {\text {for}}\quad r_{12}<\sigma(\Omega_{1},\Omega_{2},\Omega_{r})\\
		0,\,\,\quad {\text {for}}\quad r_{12}> \sigma(\Omega_{1},\Omega_{2},\Omega_{r})
	\end{array}\right..
	\label{hol_smot2.31}
\end{align}
Like the potential of interaction,  thermodynamic functions of the considered model can also be represented as the sum of two terms, the first from the hard spherocylinders and the second from the attractive interaction. In particular, the free energy of the system
\begin{equation}
	\frac{F}{N_{1}kT}=\frac{F_{0}}{N_{1}kT}+\frac{F^{\text{attr}}}{N_{1}kT},
	\label{hol_smot21}
\end{equation}
where $N_{1}$ is the total number of fluid particles, $k$ is the Boltzmann constant, $T$ is the temperature, $F_{0}$ is the contribution from hard spherocylinders, $F^{\text{attr}}$ is the contribution from the attractive part of the interaction.

To describe the contribution from hard spherocylinders, we  use the scale particle method, which was recently  developed in our group to describe a fluid of hard spheres in a porous medium \cite{29_HolDong,30_PatHolShmot2011,31_HolPatDong2013,32_HolPatDong2017} and was generalized to the case of a fluid of hard spherocylinders in a porous medium \cite{25_HolShmotPat,33_HolShmotPats2014,34_HolShmot2018}. The key point in the scale particle method for a spherocylinder system is the introduction of a scale cylinder with replaceable dimensions
\begin{equation}
	D_{s}=\lambda_{s}D_{1},\quad	L_{s}=\alpha_{s}L_{1}.
\end{equation}
The method is based on the exact calculation of the chemical potential of this large-scale particle at $\lambda_{s}\rightarrow0$ and $\alpha_{s}\rightarrow0$ and its combination with thermodynamic consideration of the large-scale particle of macroscopic size. However, the standard scaled particle theory (SPT) of the fluids in the presence of porous media contains a subtle discrepancy, which appeared in the case when the size of matrix particles is essentially larger than the size of fluid particles. This discrepancy was eliminated in \cite{30_PatHolShmot2011,31_HolPatDong2013} as part of a new approach named SPT2. The expressions, which are derived from this approach, include two parameters that define the porosity of a matrix. The first parameter is named geometrical porosity~$\phi_{0}$. It characterizes the free volume, which is not occupied by matrix particles. The second parameter is named a porosity of the probe particle $\phi$. It is defined by the chemical potential of a fluid in the limit of infinite dilution and has a meaning of probability to find a particle of fluid in an empty matrix. It was shown that the SPT2 agrees well with the data of computer simulation at low densities of fluid, while at medium and at high densities we could  obtain an essential difference. This difference becomes especially important when the packing fraction of a fluid $\eta_{1}$ reaches the values close to the porosity of the probe particle $\phi$, because the obtained expressions diverge at $\eta_{1} \rightarrow \phi$. As a result, various approximation schemes were propose. The accuracy of them was verifed by comparing theoretical results with the corresponding computer simulation data. Herein below, we  concentrate on the SPT2b1 approximation the results of which reproduce computer simulation data quite well. Omitting the results of the calculations, the details of which are given in our previous publications \cite{25_HolShmotPat, 33_HolShmotPats2014,34_HolShmot2018}, we present the final results for the chemical potential, pressure and free energy, respectively.
\begin{align}
	&\beta\left(\mu_{1}^{\text{ex}}-\mu_{1}^{0}\right)^{\text{SPT2b1}}=\sigma(f)-\ln(1-\eta_{1}/\phi_{0})+[1+A(\tau(f))]\frac{\eta_{1}/\phi_{0}}{1-\eta_{1}/\phi_{0}}
	\nonumber\\
	&+\frac{\eta_{1}(\phi_{0}-\phi)}{\phi_{0}\phi(1-\eta_{1}/\phi_{0})}
	+\frac12[A(\tau(f))+2B(\tau(f))]\frac{(\eta_{1}/\phi_{0})^{2}}{(1-\eta_{1}/\phi_{0})^{2}}\nonumber\\
	&+\frac{2}{3}B(\tau(f))\frac{(\eta_{1}/\phi_{0})^{3}}
	{(1-\eta_{1}/\phi_{0})^{3}},
	\label{hol_smot2.18}
\end{align}
\begin{align}
	&\left(\frac{\beta P}{\rho_{1}}\right)^{\text{SPT2b1}}=\frac{1}{1-\eta_{1}/\phi_{0}}\frac{\phi_{0}}{\phi}+\left(\frac{\phi_{0}}{\phi}-1\right)
	\frac{\phi_{0}}{\eta_{1}}\ln\left(1-\frac{\eta_{1}}{\phi_{0}}\right)\nonumber\\
	&+\frac{A(\tau(f))}{2}\frac{\eta_{1}/\phi_{0}}{(1-\eta_{1}/\phi_{0})^{2}}+\frac{2B(\tau(f))}{3}\frac{(\eta_{1}/\phi_{0})^{2}}{(1-\eta_{1}/\phi_{0})^{3}},
	\label{hol_smot2.19}
\end{align}
\begin{eqnarray}
	\frac{\beta F}{N}^{\text{SPT2b1}} &=&  \sigma(f)+\ln\frac{\eta_{1}}{\phi}-1
	-\ln(1-\frac{\eta_{1}}{\phi_{0}})\nonumber\\
	&+&\left(1-\frac{\phi_{0}}{\phi} \right)\bigg[1 
	+\frac{\phi_{0}}{\eta_1}\ln(1-\eta_1/\phi_{0})\bigg]  \frac{A(\tau(f))}{2} \frac{\eta_1/\phi_0}{1-\eta_1/\phi_0}\nonumber\\
	&+& \frac{B(\tau(f))}{3}
	\left(\frac{\eta_1/\phi_0}{1-\eta_1/\phi_0}\right)^2,
	\label{hol_smot2.22}
\end{eqnarray}
where
\begin{eqnarray}
	\sigma(f)=\int f(\Omega_{1})\ln f(\Omega)\rd\Omega,
	\label{hol_smot2.20}
\end{eqnarray}
\begin{equation}
	\tau(f)=\frac{4}{\piup}\int f (\Omega_{1}) f (\Omega_{2})\sin\gamma_{12}\rd\Omega_{1}\rd\Omega_{2},
	\label{hol_smot2.11}
\end{equation}
$\rd\Omega=\frac{1}{4 \piup}  \sin(\theta) \rd \theta \rd \phi $ is normalized angle element,
$f(\Omega_{1})$ is the singlet function that characterizes the orientational distribution of spherocylinders and is found by minimizing the free energy of the system.  The expressions for $A(\tau(f))$ and $B(\tau(f))$ are presented in the Appendix.

To correctly describe the thermodynamic properties according to \cite{34_HolShmot2018},  we introduce the Carnahan-Starling (CS) correction \cite{35_Carnahan69}. As a result, pressure, chemical potential and free energy are represented respectively in the forms
\begin{eqnarray}
	\frac{\beta P^{\text{SPT2b1}-\text{CS}}}{\rho_{1}}=\frac{\beta P^{\text{SPT2b1}}}{\rho_{1}}+\frac{\beta \Delta P^{\text{CS}}}{\rho_{1}},
	\label{hol_smot2.23}
\end{eqnarray}
\begin{align}
	\beta\mu_{1}^{0}=(\beta\mu_{1})^{\text{SPT2b1}}+\beta(\Delta\mu_{1})^{\text{CS}},
	\label{hol_smot2.25}
\end{align}
\begin{align}
	\frac{\beta F_{0}}{N_1}=\frac{\beta F}{N_1}^{\text{SPT2b1}}+\frac{\beta \Delta F}{N_1}^{\text{CS}},
	\label{hol_smot2.28}
\end{align}
where
\begin{eqnarray}
	\frac{\beta \Delta P^{\text{CS}}}{\rho_{1}}=-\frac{\left(\eta_{1}/\phi_0\right)^3}{\left(1-\eta_{1}/\phi_0\right)^3}.
	\label{hol_smot2.24}
\end{eqnarray}
\begin{equation}
	(\beta\Delta\mu_{1})^{\text{CS}}=
	\ln(1-\frac{\eta_{1}}{\phi_0})+\frac{\eta_{1}/\phi_0}{1-\eta_{1}/\phi_0}-\frac{1}{2}\frac{(\eta_{1}/\phi_0)^{2}}{(1-\eta_{1}/\phi_0)^{2}}-\frac{(\eta_{1}/\phi_0)^{3}}{(1-\eta_{1}/\phi_0)^{3}},
	\label{hol_smot2.27}
\end{equation}
\begin{equation}
	\left( \frac{\beta\Delta F}{N_{1}}\right) ^{\text{CS}}=
	\ln(1-\eta_{1}/\phi_0)+\frac{\eta_{1}/\phi_0}{1-\eta_{1}/\phi_0}-\frac{1}{2}\frac{\left(\eta_{1}/\phi_0\right)^2}{\left(1-\eta_{1}/\phi_0\right)^2}.
	\label{hol_smot2.29}
\end{equation}
As was discussed in \cite{27_HolShmot2020}, the van der Waals approximation corresponds to the first order of thermodynamic perturbation theory, if for the pair function of the reference system we use the approximation
\begin{equation}
	g_{2}^{0}(r,\Omega_{1},\Omega_{2})=\exp[-\beta u^{\text{rep}}(r,\Omega_{1},\Omega_{2})],
	\label{hol_smot25}
\end{equation}
which is valid in the region of low densities.
As a result, the free energy, the equation of state, and the chemical potential of the fluid can be represented in the standard van der Waals form \cite{25_HolShmotPat, 26_HolShmot2015,27_HolShmot2020}
\begin{align}
	\frac{\beta (F-F_{0})}{V}=-\rho_{1}\beta\eta_{1}a,
	\label{hol_smot2.33}
\end{align}
\begin{align}
	\frac{\beta P}{\rho_{1}}=\frac{\beta P_{0}}{\rho_{1}}-\beta\eta_{1}a,
	\label{hol_smot2.35}
\end{align}
\begin{align}
	\beta\mu_{1}=\beta\mu_{1}^{\circ}-2\beta\eta_{1}a,
	\label{hol_smot2.36}
\end{align}
where
\begin{align}
	a=-\frac{1}{2\phi_{0}V_{1}}\int f (\Omega_{1})f (\Omega_{2})u^{\text{attr}}(r,\Omega_{1},\Omega_{2}) 
	r^{2}\rd r\rd\Omega_{2}\rd\Omega_{1}\rd\Omega_{r}.
	\label{hol_smot2.34}
\end{align}
The factor $1/\phi_{0}$ excludes  the volume occupied by matrix particles at integration. For the potential we are considering (\ref{hol_smot37})
\begin{align}
	a=&-\frac{1}{2\phi_{0}V_{1}}\epsilon_{0}\left[\frac{4}{3}\piup\gamma_{1}^{3}D_{1}^{3}-\int \rd\Omega_{1}\rd\Omega_{2}f(\Omega_{1})f(\Omega_{2})V_{1}^{\text{exc}}(\Omega_{1},\Omega_{2})\right.\nonumber\\ 
	&\left.+\chi\left[\frac{4}{3}\piup\gamma_{1}^{3}D_{1}^{3}\int \rd\Omega_{1}\rd\Omega_{2}f(\Omega_{1})f(\Omega_{2})P_{2}(\cos \gamma_{12})\right.\right.\nonumber\\ 
	&\left.\left.
	-\int \rd\Omega_{1}\rd\Omega_{2}f(\Omega_{1})f(\Omega_{2})
	V_{1}^{\text{exc}}(\Omega_{1},\Omega_{2})P_{2}(\cos \gamma_{12})\right]\right],
	\label{hol_smot2.40}
\end{align}
where $\chi=\epsilon_{2}/\epsilon_{0}$,
\begin{align}
	V^{\text{exc}}_{1}\left(\Omega_{1},\Omega_{2}\right)&=\frac{1}{3}\int \rd\Omega_{r}\left[\sigma\left(\Omega_{1},\Omega_{2},\Omega_{r}\right)\right]^{3}\nonumber\\
	&=\frac{4}{3}\piup D^{3}_{1}+2\piup D^{2}_{1}L_{1}+2D_{1}L^{2}_{1}\sin\gamma_{12}\left(\Omega_{1},\Omega_{2}\right).
	\label{hol_smot2.41}
\end{align}
is the volume formed by two spherocylinders with orientations $\Omega_{1}$ and $\Omega_{2}$.
Minimization of the free energy with respect to the singlet distribution function $f(\Omega_{1})$ leads to an integral equation for the latter
\begin{align}
	&\ln f\left(\Omega_{1}\right)+\lambda+\frac{8}{\piup}C\int f\left(\Omega_{2}\right)\sin\gamma_{12}\rd\Omega_{2}\nonumber\\
	&-\frac{\eta_{1}\beta\epsilon_{0}}{\phi_{0}V_{1}}\chi\left[\frac{4}{3}\piup(D_{1}+L_{1})^{3}-\frac{4}{3}\piup D_{1}^{3}-2\piup D_{1}^{2}L_{1}\right]\int f\left(\Omega_{2}\right)P_{2}\left(\cos\gamma_{12}\right)\rd\Omega_{2}\nonumber\\
	&+\frac{\eta_{1}\beta\epsilon_{0}}{\phi_{0}V_{1}}\chi\frac{8}{\piup}2D_{1}L_{1}^{2}\int f\left(\Omega_{2}\right)\sin\gamma_{12}P_{2}\left(\cos\gamma_{12}\right)\rd\Omega_{2}=0,
	\label{hol_smot2.43}
\end{align}
where the constant $\lambda$ is determined from the normalization condition
\begin{equation}
	\int f(\Omega)\rd\Omega = 1.
	\label{hol_smot2.5}
\end{equation}
The constant $C$ is represented as a sum
\begin{equation}
	C=C_{\text{rep}}+C_{\text{attr}}\,,
	\label{hol_smot2.44}
\end{equation}
where
\begin{equation}
	C_{\text{rep}}=\frac{{\eta_{1}}/{\phi_{0}}}{1-{\eta_{1}}/{\phi_{0}}}\left[\frac{3\left(\gamma_{1}-1\right)^{2}}{3\gamma_{1}-1}\right]\left(1-\frac{P{'}_{0\lambda}}{2\phi_{0}}\right)
	+\frac{{\eta_{1}}/{\phi_{0}}}{\left(1-{\eta_{1}}/{\phi_{0}}\right)}\delta\left(\frac{6\gamma_{1}}{3\gamma_{1}-1}-\frac{P{'}_{0\lambda}}{\phi_{0}}\right),
	\label{hol_smot2.45}
\end{equation}
is the contribution from the repulsive part of the interaction. The coefficient $\delta={3}/{8}$ is due to the Parsons-Lee correction \cite{36_Parsons79,37_Lee87}, which we introduced in \cite{27_HolShmot2020}.
The contribution from the attractive part of the interaction
\begin{equation}
	C_{\text{attr}}=\frac{\eta_{1}\beta\epsilon_{0}}{\phi_{0}V_{1}}2D_{1}L_{1}^{2}.
	\label{hol_smot245}
\end{equation}
Note that in the case when $\epsilon_{2} = 0$, the equation (\ref{hol_smot2.43}) has the same structure as the corresponding equation, 
\begin{equation}
	\ln f\left(\Omega_{1}\right)+\lambda+\frac{8}{\piup}C\int f\left(\Omega_{2}\right)\sin\gamma_{12}\rd\Omega_{2}=0,
	\label{hol_smot2.47}
\end{equation}
which was obtained by Onsager \cite{21_Onsager49} for a fluid of hard spherocylinders at $L_{1} \rightarrow \infty$, $D_{1} \rightarrow 0$  and a fixed fluid concentration $c={1}/({4}\piup)\rho_{1}L_{1}^{2}D_{1}$. In this limit $C\rightarrow c$.

\section{Results and discussion}

The integral equation (\ref{hol_smot2.43}) for the singlet function $f(\Omega)$ together with the expressions (\ref{hol_smot2.35}) and~(\ref{hol_smot2.36}) for the equation of state and chemical potential is used to build phase diagrams of the system. Note that as in our previous work \cite{27_HolShmot2020}, the coexistence curves are constructed from the conditions of thermodynamic equilibrium
\begin{align}
	\mu_{1}\left(\rho_{1}^{1},T\right)&=\mu_{1}\left(\rho_{1}^{2},T\right),\nonumber\\
	P\left(\rho_{1}^{1},T\right)&=P\left(\rho_{1}^{2},T\right),
	\label{hol_smot3.1}
\end{align}
where $\mu_{1}\left(\rho_{1}^{1},T\right)$ and $P\left(\rho_{1}^{1},T\right)$ are the chemical potential and the fluid pressure, respectively, $\rho_{1}^{1} $ and $ \rho_{1}^{2} $  are the fluid densities of two different phases $1$ and $2$. The numerical solution of the equations (\ref{hol_smot3.1}) is realized by using the Newton-Rafson algorithm. Thus, in this work 
we limit ourselves to the case $\epsilon_{2}=0\,\, (\chi=0)$.

The considered fluid is characterized by three dimensionless parameters, such as the packing fraction $\eta_{1}=\rho_{1} V{_1}$, which plays the role of the density of fluid, dimensionless temperature $T^{*}=kT/\epsilon_{0}$ and parameter of asymmetry $\gamma_{1}=1+{L_{1}}/{D_{1}}$. For such a fluid in the liquid state theory \cite{22_YukhHol80}, we have the typical gas-liquid phase transition, which is described in coordinates $\eta_{1}-T^{*}$. The asymmetry of spherocylinders, which is described by parameter $\gamma_{1}$, leads to isotropic-nematic transition with shifts to smaller $\eta_{1}$ with increasing parameter $\gamma_{1}$. As a result for a small value of $\gamma_{1}$, the gas-liquid phase transition appears in an isotropic phase, while for rather big values $\gamma_{1}$ (bigger than 50), the liquid-gas transition appears in the nematic phase region \cite{19_Wu,20_Franco-Melgar}. Thus, for a bigger parameter $\gamma_{1}$ we will  have a gas-liquid transition as the transition between two nematic phases with different densities of fluid and order parameters \cite{20_Franco-Melgar,27_HolShmot2020}. Traditionally, starting from Onsager \cite{21_Onsager49}, the description of nematic phase is based on the application of trial function for the singlet distribution function $f(\Omega_{1})$ with the parameters, which can be found from the minimization of the free energy of the considered system. However, such a procedure overestimates the orientation ordering in the system \cite{23_Vroege92}. This parameter of van der Waals $a$, which is defined by equations (\ref{hol_smot2.34})--(\ref{hol_smot2.40}), can be very different for singlet distribution function, obtained from numerical solution of integral equation~(\ref{hol_smot2.43}) and for the singlet distribution function in the trial form. This effect was demonstrated by us recently in~\cite{27_HolShmot2020} where it was shown that the application of trial functions significantly overestimates the gas-liquid phase diagram in the region of high temperatures and significantly expands the density of the region of coexistence of two nematic phases. It was also shown \cite{27_HolShmot2020} that a decrease in porosity shifts the nematic-nematic transition to lower densities and lower temperatures. 


Starting to use the theory, which was given in the previous paragraph to describe the phase diagram of the PBLG solution in DMF, we note two characteristics that must be taken into account in this model. The first one is a change in the conformation of the PBLG macromolecule in the considered range of temperatures and concentrations, and the second is related to the effect of the solvent. Since the considered model is based on a hard spherocylinder, it is not clear whether we have to consider the flexibility of molecules, which would enable us to include the chain of conformation of the macromolecules. In general, the considered model should include soft enough repulsion. In such a case, similar to the theory of simple liquids \cite{37a_Barker}, we can shift our consideration to the hard spherocylinders with temperature-dependent geometry. By contrast,  in this paper we will follow the approach that was  recently proposed by Wu, Miller and Jackson \cite{19_Wu} in which the dependence from temperature of parameter of spherocylinder $\gamma_{1}(T)$ was used as the possibility for the description of phase behavior of PBLG-DMF mixture. The other aspect of considered mixture is connected with the description of solvent effects. These effects can be modelled by the dependence from temperature of attractive parameters $\epsilon_{0}(T)$ and $\epsilon_{2}(T)$. Neglecting the anisotropic attraction ($\epsilon_{2}=0$), we, as in \cite{19_Wu} within the considered model, actually have two fitting temperature-dependent parameters, which we use in this paper for theoretical reproduction of the experimentally observed phase diagram for PBLG in DMF. The first is the geometric parameter $L_{1}^{*}(T)=L_{1}/D_{1}$, which is determined by the ratio of the length of the cylinder $L_{1}$ to its diameter and takes into account changes in the conformation of the molecule with temperature. The second factor is related to the temperature dependence of the depth of the potential well $\epsilon_{0}(T)$ and is largely determined by the effect of the solvent. 

Similar to \cite{19_Wu}, we can fix the characteristic diameter $D_{1}$ of PBLG spherocylinders as 14.2 \AA,  to represent an extended low-temperature conformation, which is found to lie between the value that corresponds to the minimum intermolecular distance (13.0 \AA) 
 and the value for the backbone with extended side chains (15.2 \AA) \cite{16_Ginzburg03}. In such a way, the length of spherocylinder $L_{1}$  varies with temperature, which reflects the dependence of state of the conformations of PBLG. Alternatively, the volume of the PBLG model can be fixed so that the length and the diameter of the spherocylinder varies with temperature. From the analyses of the experimental data \cite{16_Ginzburg03} in \cite{19_Wu} for molecular anisotropy there was proposed  the following simple linear function in the inverse temperature:
\begin{equation}	
L^{*}_{\text{PBLG}}=\frac{L_{\text{PBLG}}}{D_{\text{PBLG}}}=\frac{35003.82}{T/K} -21.3179 .
\label{hol_smot4.111}
\end{equation}	
For a better description of the behavior of an isotropic-nematic phase in the present paper, this dependence was improved by the following function:
\begin{equation}
	L^{*}_{\text{PBLG}}=\frac{L_{\text{PBLG}}}{D_{\text{PBLG}}}= 13.04+\frac{813.53314}{\left(1+0.00343 T\right)^{\frac{1}{0.37313}}}.
	\label{hol_smot4.1}
\end{equation}
\begin{figure}[!t]
	\centerline{
	\includegraphics [height=0.48\textwidth]{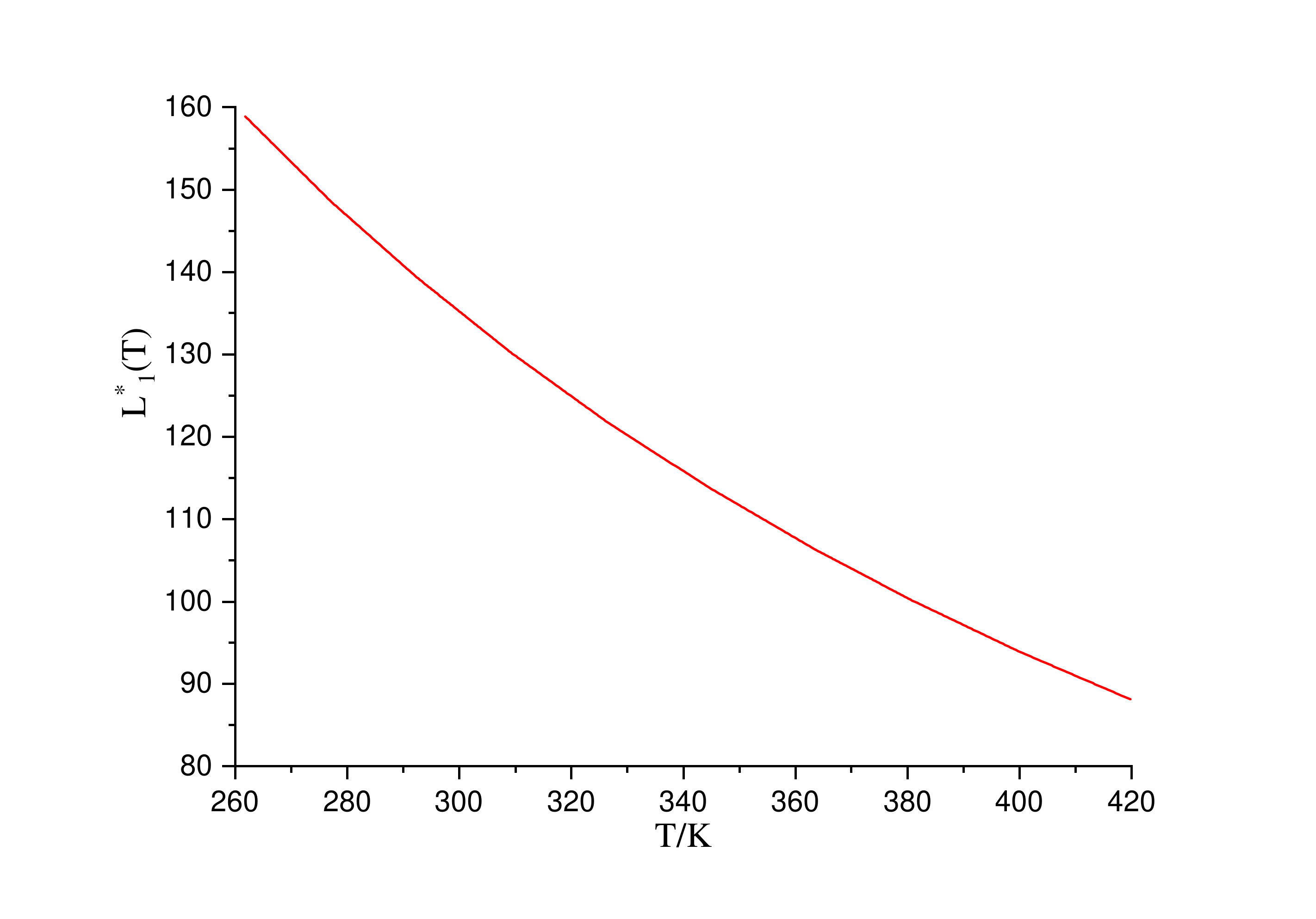}
	}
	\caption{(Colour online) Temperature dependence of the apparent length $L^*_{1}(T)$ of the PBLG spherocylinder (molecular weight $M_w = 310 000$ g mol$^{-1}$) in DMF solution, which estimated by using the generalized equation by van der Waals for anisotropic fluid (continuous curve) for the simple attractive hard-spherocylinder model.}
	\label{Fig1.1}
\end{figure}
When the temperature changes from 420 to 269~K, the effective aspect ratio is in the range from $L^{*}_{\text{PBLG}} = 62$ to $118$. We recall here that the effective diameter of the model of the spherocylinder is fixed to the value in the extended low-temperature configuration, $D_{1} = 14.2$ \AA. Thus, the apparent length of the \text{PBLG} spherocylinder ranges from $L_1 \sim 88$ to $168$ nm (see figure~\ref{Fig1.1}), which is generally consistent with the range of the lengths of stability, that were obtained experimentally \cite{16_Ginzburg03} from the light scattering studies, such as $L_{\text{PBLG}}^{\text{exp}} \sim 70$ up to $160$ nm, diameter $D_{\text{PBLG}}^{\text{exp}} \sim 15.2\,\,\text{\AA}$ estimated experimentally \cite{16_Ginzburg03}, when the side chains of the PBLG spherocylinder are independent rotators.  The obtained results are also consistent for the case when the volume of the spherocylinder $V_{\text{PBLG}}$ is fixed. In this case, the diameters are in the range from $D_{\text{PBLG}} = 14.2\,\,\text{\AA}$ at 260 K to 18.0~\AA\,at 420~K, because it is the best reflection of the experimental phase behavior, which corresponds to an average of 16.1 \AA. The temperature dependence of the depth of the potential well $\epsilon_{0}(T)$ is shown in figure~\ref{Fig1.2} and can be represented by the following function:
%
\begin{equation}
	\epsilon_{0}(T)=0.2168+\frac{0.044}{1+\exp \left( \frac{T-285.7959}{1.6964}\right)  }.
	\label{hol_smot4.3}
\end{equation}
\begin{figure}[!t]
	\centerline{
		\includegraphics [height=0.5\textwidth]{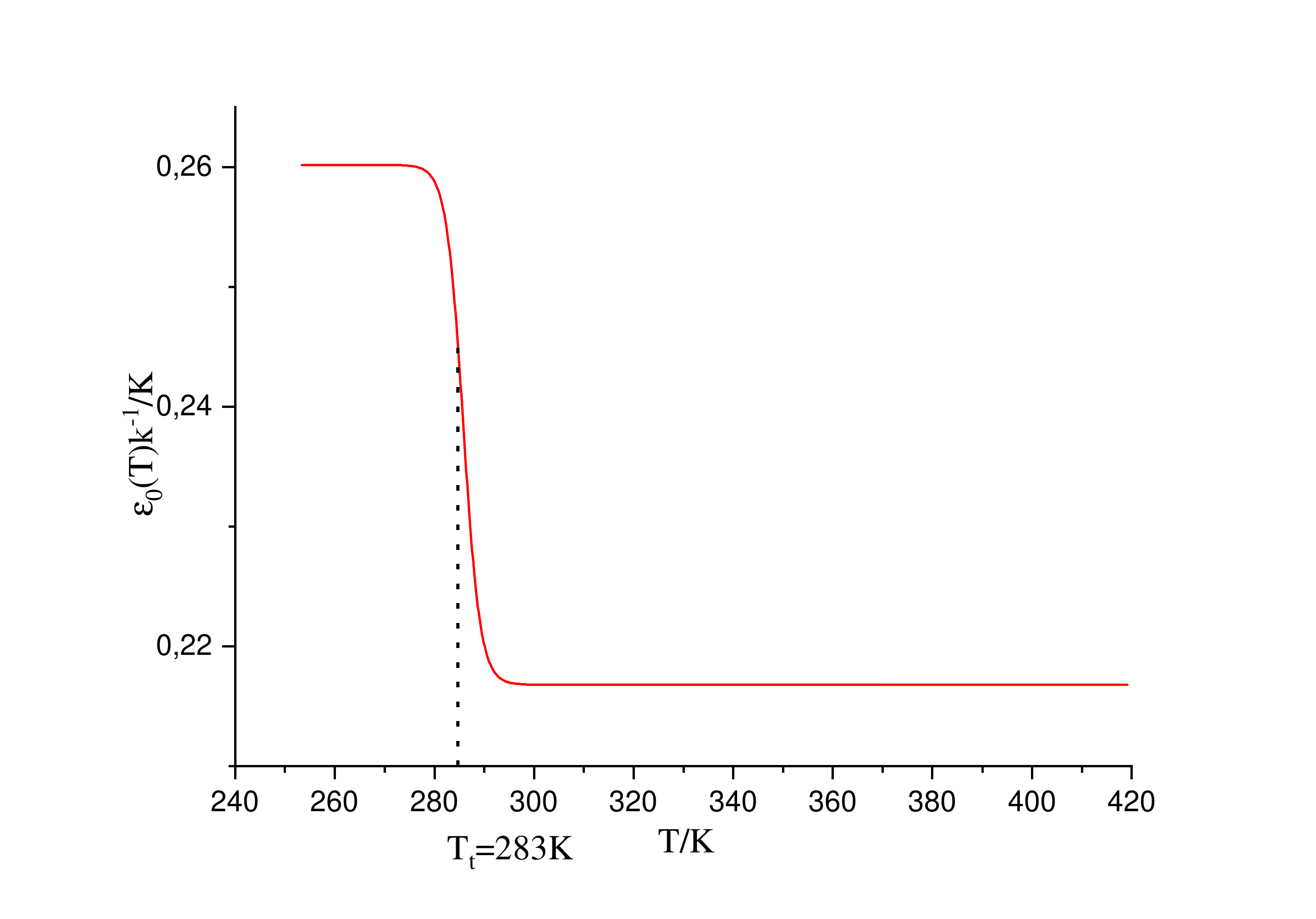}
	}
	\caption{(Colour online) Temperature dependence of isotropic attractive parameter $\epsilon_{0}(T)$ for solutions of PBLG, $k$ is Boltszman constant (molecular weight $M_w = 310 000$ g mol$^{-1}$) in DMF, which is estimated by using the generalized equation by van der
		Waals for anisotropic fluid (continuous curve) for the simple attractive hard-spherocylinder model. The triple temperature $T_t$ is indicated.}
	\label{Fig1.2}
\end{figure}
\begin{figure}[!t]
	\centerline{
		\includegraphics [height=0.5\textwidth]{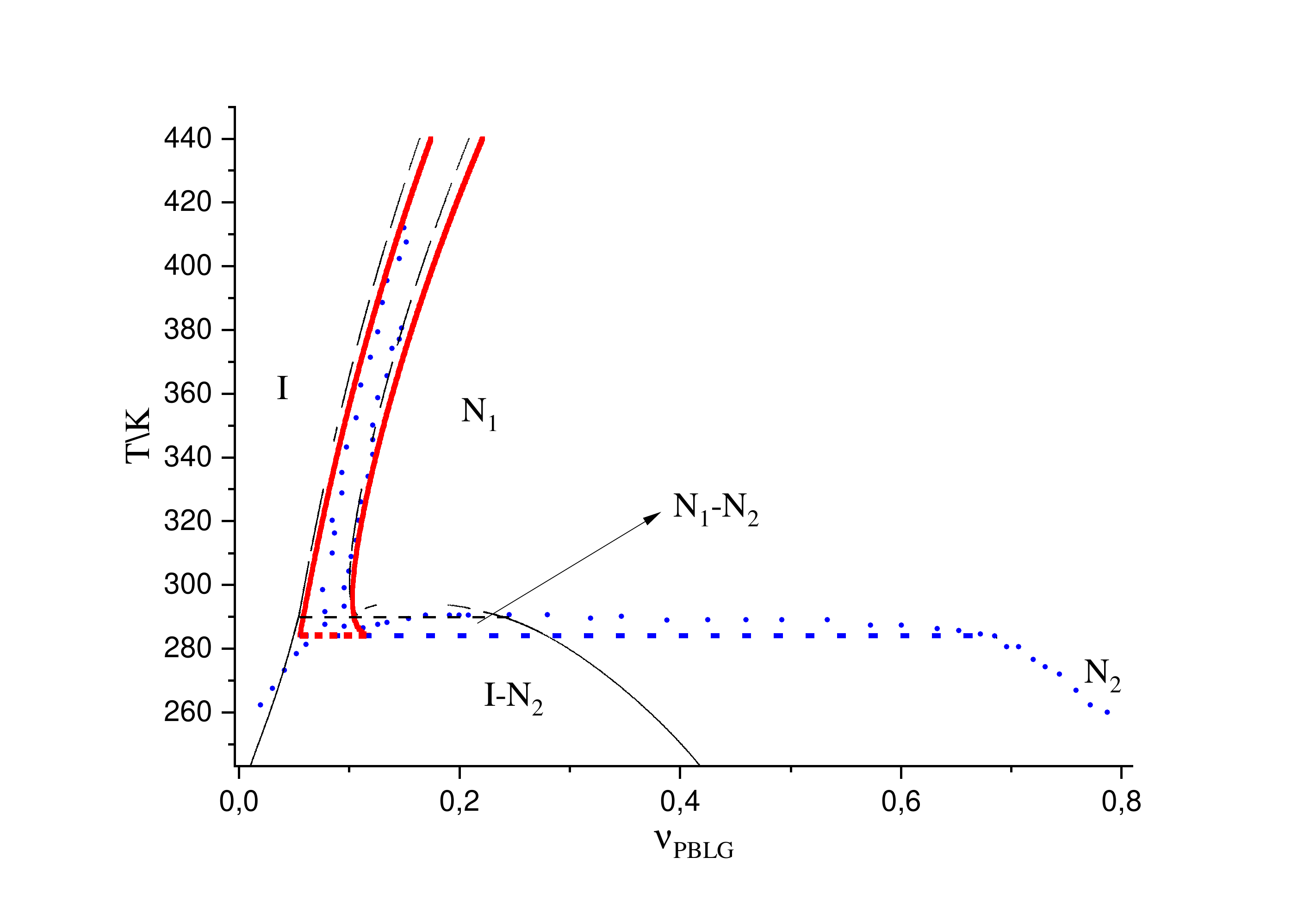}
	}
	\caption{(Colour online) Temperature T --- volume fraction $\nu_{\text{PBLG}}$ representation of the phase
		diagram for solutions of the polypeptide PBLG (molecular weight $M_w = 310 000$ g mol$^{-1}$) in DMF. The symbols represent the experimental data by Miller and
		co-workers \cite{10_WeeMil1971,11_MilWuWeeSanRaiGo74,12_ZimWuMilMas83,13_RusMil1983,14_RusMil1984}. The continuous solid curves correspond to the theoretical description, which is obtained by using the generalized equation by van
		der Waals for the attractive model of hard spherocylinder. The red curves correspond to the description of apparent length $L^*_{1}(T)$ of PBLG molecules in the
		form (\ref{hol_smot4.111}), and  thin black curves correspond to the description of $L^*_{1}(T)$ in
		the form (\ref{hol_smot4.1}). The two regions of isotropic-nematic coexistence correspond to
		$I-N_1$ and $I-N_2$, the region of nematic-nematic coexistence corresponds to $N_1-N_2$, and
		the dot-dashed line denotes the $I-N_1$, $I-N_2$ --- three-phase coexistence.}
	\label{Fig1.3}
\end{figure}
This dependence is very close to the corresponding dependence, which is proposed in \cite{19_Wu}. As we can see from figure~\ref{Fig1.3}, the change of the depth of the potential well is quite small in the temperature range ($\Delta\epsilon_{0}/k \sim 0.04$~K) and does not affect the phase behavior for gas-like states at a low concentration. In the range of temperatures below 280~K, the parameter of energy $\epsilon_{0}(T<280$ $\text{K})/k \simeq 0.26$~K, while the parameter takes a value of $\epsilon_{0}(T>290$ $\text{K})/k \simeq 0.22$~K. In other words, the isotropic-nematic phase behavior, which is detected by the PBLG-DMF system, can be essentially interpreted as two separate lyotropic liquid crystal transitions on both sides of the triple point. However, this is not the case for the nematic-nematic region of coexistence between two anisotropic phases. The effective attractive interactions between PBLG spherocylinders are stronger (correspond to higher values of $\epsilon_{0}$) in states with high concentration (liquid-like) and with low temperature than in states with low concentration (gas-like) and with high temperature, which indicate a significant interaction of molecules with solvent (e.g.., dielectric constant for electrolytes).
The parametrization (\ref{hol_smot4.111}) or (\ref{hol_smot4.1}) for the geometrical parameter of PBLG molecules and (\ref{hol_smot4.3}) for the depth of effective attraction between PBLG molecules $\epsilon_{0}(T)$ characterizes the influence of the change of conformation of molecules with the change of temperature on the geometry of molecules and the influence of the solvent on the attraction between the molecules, correspondingly. We see that with decreasing temperature, PBLG molecules are more stiff and the effective length of molecules increases. Accordingly, the role of the solvent in the attraction between the molecules also increases for lower temperatures and higher concentrations.

The parameters $\epsilon_{0}(T)$ and $L_{\text{PBLG}}^{*}(T)$ with temperature dependence, which is given by the expressions (\ref{hol_smot4.3}) and (\ref{hol_smot4.1}), were used to construct a PBLG phase diagram in DMF. In this case, as in \cite{19_Wu}, instead of the packing parameter $\eta_{1}$ of the stiff polymer, we used the parameter of the volume fraction of polymer $V_{\text{PBLG}}$, which is linked to the parameter $\eta_{1}=\eta_{\text{PBLG}}$ by a simple relation
\begin{equation}
	V_{\text{PBLG}}=\frac{\eta_{\text{PBLG}}M_{w}}{N_{\text{A}}\rho_{\text{PBLG}}V_{m,\text{PBLG}}},
	\label{hol_smot41}
\end{equation}
where $N_{\text{A}}$ is the Avogadro's number, $V_{m,\text{PBLG}}$ is the volume taken up by the molecule
\begin{equation}
	V_{m,\text{PBLG}}=\piup\frac{D_{\text{PBLG}}^{3}}{6}+\frac{1}{4}\piup L_{\text{PBLG}}D_{\text{PBLG}}^{2}=\frac{\piup}{2}D_{\text{PBLG}}^{3}\left(\frac{1}{3}+\frac{1}{2}L_{\text{PBLG}}^{*}\right).
	\label{hol_smot42}
\end{equation}
Since the molecular weight $M_{w}$ and the density $\rho_{\text{PBLG}}$ depend on the polymerization conditions \cite{38_LonRus1991,39_InShNos2000}, we  take for them the values $M_w = 310 000$ g mol$^{-1}$, $ \rho_{\text{PBLG}}=1.283$ g cm$^{-3}$.

\begin{figure}[!b]
	\centerline{
		\includegraphics [height=0.5\textwidth]{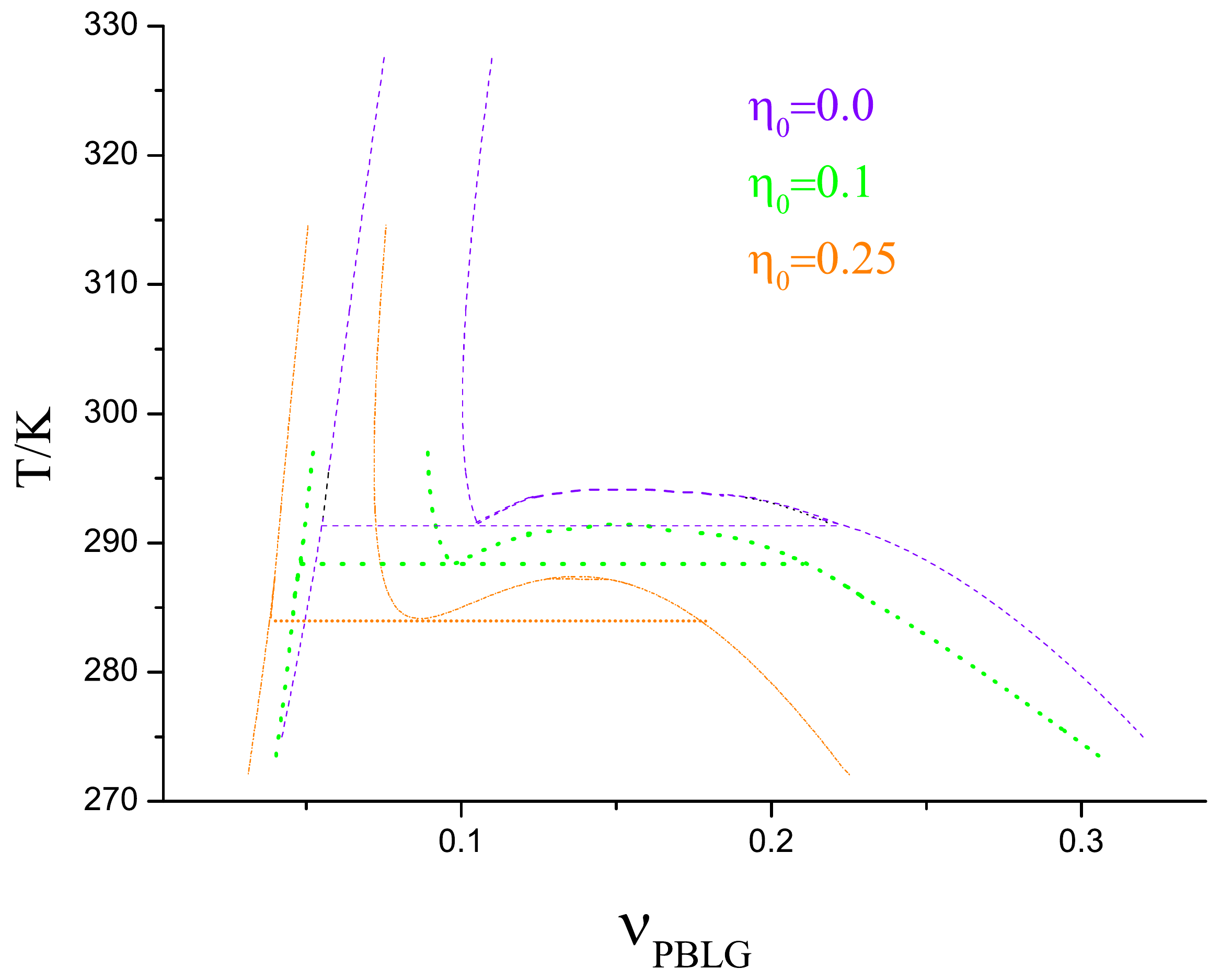}
	}
	\caption{(Colour online) Phase diagrams for a hard fluid of spherocylinder with anisotropic
		attraction in a disordered porous medium, which is calculated from the generalized equation by van der Waals for anisotropic fluid in disordered porous media.}
	\label{Fig1.4}
\end{figure}

The calculated phase diagram for a solution of PBLG in DMF for the bulk case and comparison with the corresponding experimental data are presented in figure~\ref{Fig1.3}. As you can see, the model used successfully reproduces the phase behaviour. In particular, isotropic-nematic three-phase coexistence is predicted at the temperature $T_{\text{tr}}=283$~K, which is slightly underestimated compared to the experimental value of $T_{\text{tr}}^{\text{exp}}=285.4$~K. At this point, the isotropic phase, the low-density nematic phase, and the high-density nematic phase coexist simultaneously. In figure~\ref{Fig1.3} we also show the phase diagram of PBLG in DMF, for the calculation of which a simpler formula (\ref{hol_smot4.1}) is used~\cite{19_Wu}  
As we can see, the use of this formula leads to a slightly inflated value of the tricritical point $T_{\text{t}}=291$~K. The relative simplicity of the dependence~(\ref{hol_smot4.111}) also allowed us to obtain the region of coexistence of two nematic phases, which, unfortunately, is underestimated nearly two-fold compared to the experimental results. We hope to improve the description of the phase diagram in the region of coexistence of two nematic phases by extending the model and introducing the anisotropic part of the attraction with the third fitting parameter~$\epsilon_{2}(T)$. As we showed earlier \cite{26_HolShmot2015}, the inclusion of an anisotropic attraction can expand the region of coexistence of two nematic phases. The other fit parameter $\epsilon_{0}(T)$ must, of course, also be modified.

The effect of the porous medium on the phase diagram is presented in figure~\ref{Fig1.4}. As can be seen from this figure, the porous medium shifts the phase diagram to the region of lower densities and lower temperatures. When constructing phase diagrams, we used the temperature dependence in the form (\ref{hol_smot4.111}) for the geometric parameter $L_{\text{PBLG}}^{*}$ to facilitate calculations. As can be seen from the figure, the tricritical temperature varies from  $T_{\text{tr}}=291$~K in the bulk case ($\eta_{0}=0$) to 288 K for $\eta_{0}=0.1$ and to 284~K for $\eta_{0}=0.25$, where $\eta_{0}$ is the matrix packing fraction.

\begin{figure}[!t]
	\centerline{
		\includegraphics [height=0.47\textwidth]{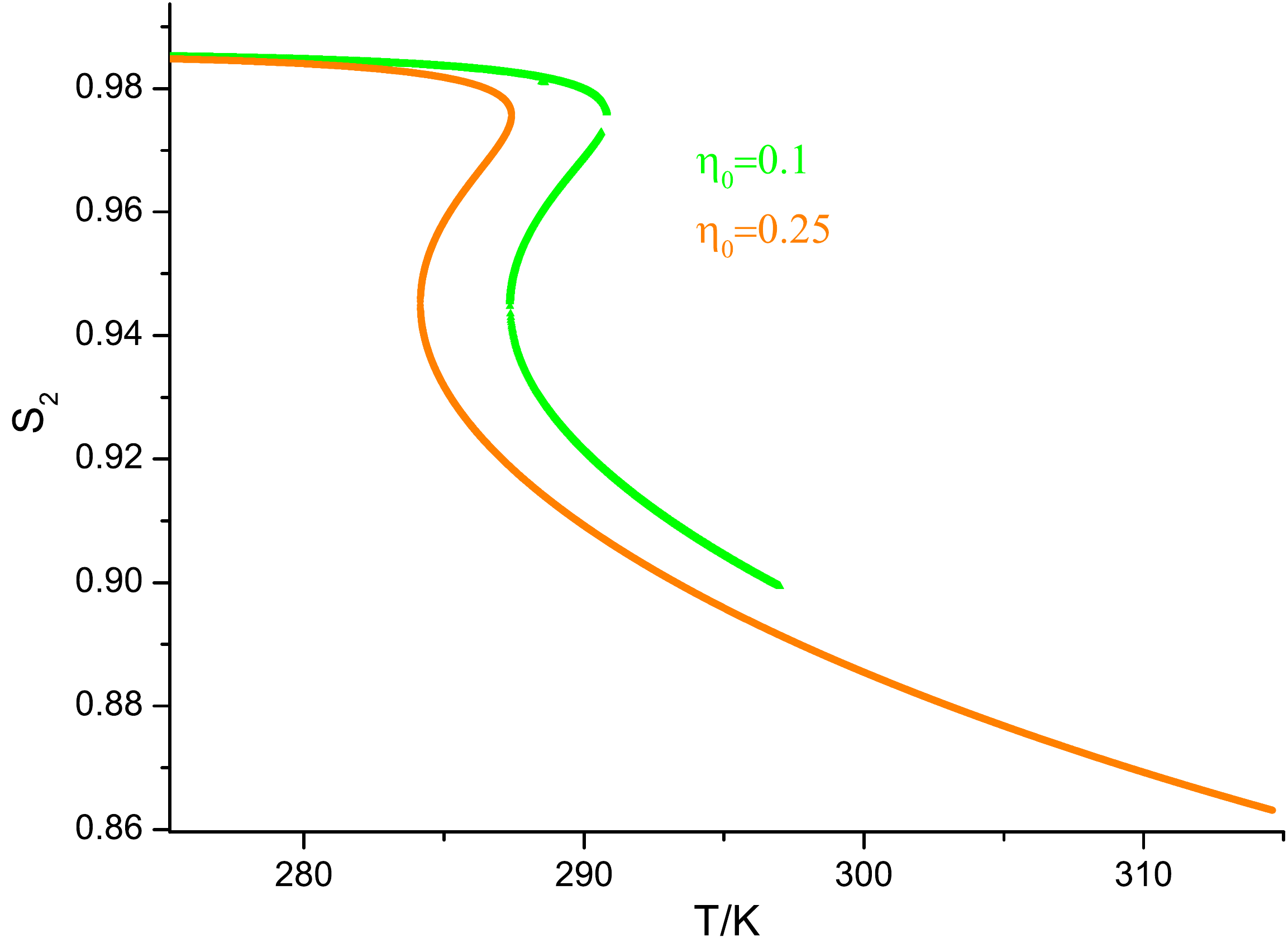}
	}
	\caption{(Colour online) Temperature dependence of the order parameter in the coexistence
		of nematic phases for a hard fluid of spherocylinder with an anisotropic attraction in a disordered porous medium, which is calculated from the generalized
		equation by van der Waals for anisotropic fluid in disordered porous media.}
	\label{Fig1.5}
\end{figure}
Finally,  figure~\ref{Fig1.5} shows the temperature dependence of the nematic order parameter
\begin{equation}
	S_{2}=\int \rd\Omega f(\Omega)P_{2}(\cos\vartheta),
	\label{hol_smot44}
\end{equation}
where $P_{2}(\cos\vartheta)=\frac{1}{2}(3\cos^{2}\vartheta-1)$ is the second order Legendre polynomial.

As can be seen from the figure, the nematic order parameter of the dense phase $N_{2}$ is close to unity and slightly decreases with an increasing temperature, whereas in the low-density phase $N_{1}$, the order parameter decreases monotonously with an increasing temperature. The porous medium also reduces the value of the order parameter.

Thus, in this paper, within the framework of the generalized van der Waals equation for hard spherocylinders with anisotropic attraction in the form of a potential well, we were able to reproduce the main features of the PBLG phase diagram in DMF by introducing the temperature dependences of the depth of the potential well $\epsilon_{0}(T)$ and of the geometric parameter $L_{\text{PBLG}}^{*}$. These features include the existence of two nematic phases which, as we have shown, are the result of competition between the attractive part of the interaction and the excluded volume of macromolecules. We have shown that the region of coexistence of two nematic phases is confined between the triple point of the isotropic phase and the two nematic phases and the critical point of coexistence of two nematic phases. The presence of a porous medium shifts the phase diagram to the region of lower densities and lower temperatures.


\section*{Acknowledgement}

Myroslav Holovko and Volodymyr Shmotolokha gratefully acknowledge financial support from the National Research Foundation of Ukraine (project No. 2020.02/0317).
\section*{{Appendix}}

\begin{eqnarray}
	A(\tau(f)) &=&6+\frac{6\left(\gamma_{1}-1\right)^2\tau(f)}{3\gamma_{1}-1}-
	\frac{p'_{0\lambda}}{\phi_0}\left(4+\frac{3\left(\gamma_{1}-1\right)^2\tau(f)}{3\gamma_{1}-1}\right) \nonumber\\
	&-&\frac{p'_{0\alpha}}{\phi_0}\left(1+\frac{6\gamma_{1}}{3\gamma_{1}-1}\right)-\frac{p''_{0\alpha\lambda}}{\phi_0} \nonumber\\
	&-&\frac{1}{2}
	\frac{p''_{0\lambda\lambda}}{\phi_0}+2\frac{p'_{0\alpha}p'_{0\lambda}}{\phi_0^{2}}+\left(\frac{p'_{0\lambda}}{\phi_0}\right)^2,
	\label{hol_smot2.9}
\end{eqnarray}
\begin{eqnarray}
	B(\tau(f)) &=&\left(\frac{6\gamma_{1}}{3\gamma_{1}-1}-\frac{p'_{0\lambda}}{\phi_0}\right) \nonumber\\  &&\times\left(\frac{3\left(2\gamma_{1}-1\right)}{3\gamma_{1}-1}+
	\frac{3\left( \gamma_{1}-1\right)^2\tau(f)}{3\gamma_{1}-1}
	-\frac{p'_{0\alpha}}{\phi_0}
	-\frac{1}{2}\frac{p'_{0\lambda}}{\phi_0}\right),
	\label{hol_smot2.10}
\end{eqnarray}
$p{'}_{0\lambda}, p{'}_{0\alpha}, p{''}_{0\alpha\lambda}$ and $p_{0\lambda\lambda}$ are corresponding derivatives of the function, $p_{0}(\alpha_{s},\lambda_{s})$ at $\alpha_{s}=\lambda_{s}=0$. Function
\begin{equation}
	p_{0}(\alpha_{s},\lambda_{s})=\exp[-\beta\mu_{s}^{0}(\alpha_{s},\lambda_{s})]
	\label{hol_smot22}
\end{equation} 
has the meaning of the probability of finding in a cavity the system which was created by a large-scale particle in the absence of fluid particles and is determined by the excess of the chemical potential $\mu_{s}^{0}(\alpha_{s},\lambda_{s})$ of a large particle at an infinite dilution of the fluid. The function $p_{0}(\alpha_{s},\lambda_{s})$ also introduces two important characteristics of the porous medium, namely the geometric porosity
\begin{equation}
	\phi_{0}=p_{0}(\alpha_{s}=0, \lambda_{s}=0),
	\label{hol_smot23}
\end{equation}
and the thermodynamic porosity
\begin{equation}
	\phi=p_{0}(\alpha_{s}=1, \lambda_{s}=1)=\exp(-\beta\mu_{1}^{0}),
	\label{hol_smot24}
\end{equation}
which are determined by the excess of the chemical potential of the fluid at an infinite dilution $\mu_{1}^{0}$.

\newpage

\ukrainianpart

\title{Вплив пористого середовища на фазову поведінку розчинів поліпептидів}
\author{В. І. Шмотолоха, М. Ф. Головко}
\address{
Інститут фізики конденсованих систем НАН України, вул. Свенціцького, 1, 79011 Львів, Україна
}

\makeukrtitle

\begin{abstract}
\tolerance=3000%
 Запропоноване в попереднiх роботах авторiв узагальнене рiвняння ван дер Ваальса для анiзотропних плинiв у пористих
середовищах використано для опису впливу пористих середовищ на
фазову поведiнку розчинiв полiпептидiв. Шляхом введення температурної залежностi для глибини потенцiальної ями i геометричних
параметрiв сфероцилiндра вiдтворено основнi риси фазової поведiнки полiпептиду полi ($\gamma$-бензил-$L$-глутамату) (PBLG) в розчинi диметилформамiду, включаючи iснування двох нематичних фаз. Показано, що наявнiсть пористого середовища зсуває фазову дiаграму
в область менших густин i нижчих температур.

\end{abstract}

\end{document}